\documentclass[12pt,nohyper,notoc]{JHEP}

\usepackage{amsmath,amssymb,cite,graphicx} % use drftcite in draftmode otherwise
	
\font\zfont = cmss10 %scaled \magstep1
\newcommand\ZZ{\hbox{\zfont Z\kern-.4emZ}}
\def\inbar{\vrule height1.5ex width.4pt depth0pt}
\def\IC{\relax\hbox{\kern.25em$\inbar\kern-.3em{\rm C}$}}

\newcommand{\EQ}[1]{\begin{equation} #1 \end{equation}}

\title{Black Diamonds at Brane Junctions}

\author{Andrew Chamblin$^a$, Csaba Cs\'aki$^{b,}$\footnote{J. Robert Oppenheimer Fellow}, 
Joshua Erlich$^b$ and Timothy J. Hollowood$^{b,c}$\\
$^a$CTP, MIT Bldg. 6-304, 77 Massachusetts Ave, Cambridge, MA 02139, USA\\
$^b$Theory Division T-8, Los Alamos National Laboratory, Los Alamos,
NM 87545, USA\\
$^c$Department of Physics, University of Wales Swansea,
Swansea, SA2 8PP, UK\\

Email: {\tt chamblin@ctpblack.mit.edu,  csaki@lanl.gov, erlich@lanl.gov, pyth@skye.lanl.gov}}

\abstract{We discuss the properties of black holes in brane-world 
scenarios where our universe is viewed as a four-dimensional
sub-manifold of some higher-dimensional spacetime. We consider in
detail such a model where four-dimensional spacetime lies at the
junction of several domain walls 
in a higher dimensional anti-de Sitter spacetime.  In this model
there may be any number $p$ of infinitely large extra dimensions transverse to
the brane-world.  We present an exact solution describing a black
$p$-brane which will induce on the brane-world the Schwarzschild
solution.  This exact solution is unstable to the Gregory-Laflamme
instability, whereby long-wavelength perturbations
cause the extended horizon to fragment.  We therefore argue that
at late times a non-rotating uncharged black hole in
the brane-world is described by a deformed
event horizon in $p+4$ dimensions which will induce,
to good approximation, 
the Schwarzschild solution in the four-dimensional brane world.  
When $p=2$, this deformed horizon resembles
a black diamond and more generally for $p>2$, a polyhedron.}

\preprint{{\tt hep-th/0002076}}

\begin{document}

Motivated by phenomenological considerations, 
there has recently been an enormous amount of interest
in the possibility that there may exist extra dimensions of space 
which are quite large.  In this framework the 
universe would be a brane embedded in some higher dimensional
spacetime.  In particular, this is the basic assumption underlying the models of
Randall and Sundrum \cite{RS,RS2}.
Their second model involves a thin ``distributional" static flat
domain wall, or three-brane,
separating two regions of five-dimensional anti-de-Sitter (AdS) spacetime.
They solve for the linearized graviton perturbations and find a 
square integrable bound state representing a gravitational
wave confined to the domain wall. They also find the linearized
bulk or ``Kaluza-Klein" graviton modes, and argue that
they decouple from the brane and 
make negligible contributions to the gravitational force between two
sources in the brane, so that this force is due primarily to the 
bound state.  In this way they recover an inverse square law
attraction rather than the inverse cube law one might 
na\"\i vely have anticipated in five dimensions.  
Thus, their work demonstrates that it is
possible to localize gravity on a three-brane 
when there is just one infinitely large extra dimension of space, so that the
three-brane is a domain wall.
There are many papers which serve as background for this
work; for a comprehensive list of references see \cite{US}.

The localization of gravity on a domain wall is rather striking and raises 
various questions.
For example, is it possible to localize gravity on a brane or brane 
intersection when there
is {\it more} than one large extra dimension of space?  In fact,
Arkani-Hamed {\it et al.\/}~\cite{ADDK} have shown in a simple
generalization of the Randall-Sundrum scenario that this is indeed
possible (see also \cite{CS,Ann,SeanMark,Nam}).
Their model may be outlined as follows:  
since it is possible to localize gravity on a domain wall in AdS, 
consider $p$ different domain walls (each with world-volume space dimension $d-2$)
in a $d=(p+4)$-dimensional background spacetime.  These branes can intersect
at a four-dimensional junction.  If the bulk spacetime between
the branes consists of $2^p$ patches of $(p+4)$-dimensional AdS, then it 
turns out that on the four-dimensional intersection there is
a normalizable graviton mode and so four-dimensional gravity is localized
on the brane junction.
This intersecting brane scenario is not the only possible way to
localize gravity in higher dimensions. 
For example, four-dimensional gravity can also be localized on a three-brane in
higher dimensions \cite{CK,Erich,Nima2,Nelson2,US} although in this case the relevant
geometry is not AdS. 

If matter trapped on a brane undergoes gravitational collapse then a
black hole will form. Such a black hole will have a horizon that
extends into the dimensions transverse to the brane: it will be a
genuinely $d$-dimensional object.  
Within the context of any brane-world scenario, we need to make
sure that the metric on the brane-world, which is induced by the 
higher-dimensional metric describing the gravitational collapse, is simply
the Schwarzschild solution, up to some corrections that are
negligible small so as to be consistent with current observations. In
this way we 
shall recover the usual astrophysical properties of black holes and stars
({\it e.g.\/}~perihelion precession, light bending, {\it etc\/}.).
The study of the problem of gravitational collapse in the second
Randall-Sundrum model was initiated in a recent paper \cite{CHR} (see
also \cite{EHM,GKR}).  There,
the authors proposed that what would appear to be a four-dimensional black hole
from the point of view of an observer in the brane-world, is really
a five-dimensional ``black cigar'' which extends into the extra fifth
dimension (or 
more accurately a ``black cigar butt'', or equivalently a ``pancake'' \cite{EHM,GKR},
because the object only extends a small proper distance
in the transverse space compared with its extent on the brane). 
If this cigar were to extend all the way down to the AdS horizon,
then we would recover the metric for a black string in AdS.  However,
such a black string is unstable near the AdS
horizon.  This instability, known as the ``Gregory-Laflamme'' instability,
basically means that the string will want to fragment in the region near
the AdS horizon.  However, the solution is stable far from the AdS
horizon near the domain wall.  Thus, these authors concluded that the late time solution 
describes an object which looks like the black string
far from the AdS horizon (so the metric on the domain wall is
approximately Schwarzschild) but has a horizon that closes off before reaching the
AdS horizon forming the ``tip'' of the black cigar. 
In the analogous situation in one dimension lower (where the domain
wall localizes three-dimensional gravity in a larger four-dimensional
$AdS_4$ background) the exact metric describing the situation is known
\cite{EHM} since it is an example of the C-metric in
$AdS_4$. Unfortunately, the generalization of this metric to
$AdS_{p+4}$ is not known at present and so we have to proceed in a more intuitive
fashion to arrive at a consistent picture.

In this paper, we extend the analysis of gravitational collapse on the 
brane-world to general situations. In particular, for the purposes of
illustration we will consider in detail
the model where the brane-world universe
is actually a brane-junction \cite{ADDK}: a region where multiple domain walls
intersect. However, our formalism can easily be used to describe
smoothings of the original
Randall-Sundrum scenario, where the three-brane is smeared in the
extra dimension \cite{cveticdw,DFGK,Chamblin,Gremm,US} and also other higher-dimensional
situations; for instance a three-brane embedded in $d>5$ dimensions \cite{CK,Erich,US}.
As we proceed we shall work as far as possible in a general framework
with a $d$-dimensional background which
has a four-dimensional Poincar\'e symmetry (the restriction to four dimensions
is unnecessary, but is the case most relevant for phenomenology):
\EQ{
ds^2\equiv g_{\mu\nu}(z)dx^\mu\,dx^\nu=e^{-A(z)}\eta_{ab}\,dx^a\,dx^b+g_{ij}(z)dz^i\,dz^j\ .
\label{metric}
}
Here, $x^\mu=(x^a,z^i)$, where $x^a$, for $a=0,\ldots,3$, 
are the usual coordinates of four-dimensional Minkowski space
and $z^i=x^{i+3}$, for $i=1,\ldots,p$, are the coordinates on the $p=(d-4)$-dimensional
transverse space.\footnote{In our conventions the metric $g_{\mu\nu}$
has signature $(-,+,+,+,\ldots)$.} The metric \eqref{metric} has the form of a
``warped product'' which under certain conditions is responsible for the
localization of gravity. In the general case, the localization of
gravity in four-dimensions depends on the normalizability, or otherwise, of a
certain fluctuating mode in the transverse space representing the
four-dimensional graviton. The explicit
requirement for normalizability just depends upon the metric and is \cite{US}
\EQ{
\int d^pz\,g^{00}(z)\sqrt{g(z)}<\infty\ ,
}
where $g(z)=|{\rm det}\,g_{\mu\nu}(z)|$. In \cite{US}, it was further shown that under
very general conditions the normalizability of the
four-dimensional mode also implies the decoupling of the associated
Kaluza-Klein modes.

In particular, for purposes of illustration, we will be interested in a 
simple example of the general situation described by \eqref{metric}
corresponding to the intersecting brane scenario of \cite{ADDK}. In this
example, we glue $2^p$ patches of $AdS_{p+4}$ together along  
$p$ surfaces which play the r\^ole of 
$p$ domain walls. Once we have performed this cutting and pasting,
the metric of the multi-dimensional patched AdS spacetime which describes
the domain wall junction can be written \cite{ADDK}:
\EQ{
ds^2=\frac1{\big(1+k\sum_{i=1}^{p}|z^i|/\sqrt{p}\big)^2}
\big(\eta_{ab}\,dx^a\,dx^b+dz^idz^i\big)\ ,
\label{mdrsmetric}
}
where we have included the factor of $\sqrt p$ for convenience so that
$k^{-1}$ is the conventional length scale of the AdS bulk.
%\footnote{The length
%scale $k^{-1}$ is related to the bulk cosmological constant $\Lambda$ and
%Newton constant ${\kappa}^2$ in the usual way:
%$k^2=2{\kappa}^{2}\Lambda/(p+3)(p+2)$.} 
This metric represents $p=d-4$ intersecting $(d-1)$-branes, located at
$z^i=0$, for each $i$, which mutually intersect in a four-dimensional junction located at $z^i=0$.

We want to know how to describe the endpoint of gravitational collapse
in our four-dimensional world.  
Following the work of \cite{CHR}, a natural guess is simply to replace
the (3+1) Minkowski metric appearing in \eqref{metric} 
with the (3+1) Schwarzschild metric.
Indeed, shortly, we shall prove that the following metric is still a solution
of the bulk Einstein equations with the requisite source term:
\EQ{
ds^2=e^{-A(z)}
\big(-U(r)dt^2+U(r)^{-1}dr^2+r^2(d\theta^2+\sin^2\theta d\phi^2)\big)
+g_{ij}(z)dz^i\,dz^j\ .
\label{emsm}
}
where $U(r)=1-2G_4M_4/r$.  Clearly, this metric describes a black hole horizon which
is extended in $p$ extra dimensions.  In other words, this is the metric of a
``black $p$-brane'' in the higher-dimensional spacetime.  Na\"\i
vely, it follows that
if we wish to describe a black hole on the four-dimensional brane-world all we
have to do is replace the Minkowski metric $\eta_{ab}$ in \eqref{metric} with
the four-dimensional Schwarzschild metric.

We now show that we can generalize the metric \eqref{metric} to
\EQ{
ds^2\equiv g_{\mu\nu}dx^\mu\,dx^\nu=e^{-A(z)}\tilde g_{ab}(x)\,dx^a\,dx^b+g_{ij}(z)dz^i\,dz^j\ ,
\label{gmetric}
}
where $\tilde g_{ab}(x)$ is {\em any\/} four-dimensional Ricci flat metric,
and still satisfy Einstein's equations.
First of all, let us compute
the change in the Einstein tensor when we replace the four-dimensional
Minkowski metric by the Ricci-flat metric: $\eta_{ab}\to\tilde g_{ab}$. 
It is convenient to use the the
general form of the Einstein tensor for metrics of the form $g_{\mu\nu}=e^{-A}\tilde{g}
_{\mu\nu}$ (see for example \cite{Wald}): 
\EQ{
G_{\mu\nu}=\widetilde{G}_{\mu\nu}+\tfrac{d-2}{2}\Big[\tfrac12
\widetilde{\nabla}_\mu A\,
\widetilde{\nabla}_\nu A+\widetilde{\nabla}_\mu\widetilde{\nabla}_\nu
A-\tilde{g}_{\mu\nu}\big(\widetilde{\nabla}_\rho\widetilde{\nabla}^\rho
A
-\tfrac{d-3}{4}\,\widetilde{\nabla}_\rho A\,\widetilde{\nabla}^\rho
 A\big)\Big]\ .
\label{conformal}
}
The new metric
\EQ{
d\tilde s^2\equiv\tilde g_{\mu\nu}\,dx^\mu\,dx^\nu=
\tilde g_{ab}(x)\,dx^a\,dx^b+e^{A(z)}g_{ij}(z)\,dz^i\,dz^j\ .
\label{cmetric}
}
now represents a genuine (unwarped) product of a four-dimensional space with metric
$\tilde g_{ab}(x)$ and $p$-dimensional space with metric $\tilde g_{ij}(z)=e^{A(z)}g_{ij}(z)$.
Using \eqref{conformal}, it is easy to see that the only 
components of the Einstein tensor which change when we replace
$\eta_{ab}\to\tilde g_{ab}$ are $G_{ab}$. Using the fact that
$\tilde g_{ab}$ is Ricci flat, so $\tilde G_{ab}(\tilde
g_{ab})=0$, the change is easily computed:\footnote{Notice that since
$G_{ab}^{(0)}\propto\eta_{ab}$ and $T_{ab}^{(0)}\propto\eta_{ab}$ 
the following expressions in \eqref{trgg} and \eqref{trst} are
symmetric in $a$ and $b$.}
\EQ{
\Delta G_{ab}=G_{\phantom{(0)}a}^{(0)\ c}\Delta\tilde g_{cb}\ ,
\label{trgg}
}
where $\Delta\tilde g_{ab}=\tilde g_{ab}-\eta_{ab}$ and
$G_{\mu\nu}^{(0)}$ is the original Einstein tensor for the metric
\eqref{metric}. The remaining part of the proof requires us to show
that the stress-tensor changes in a similar fashion: {\it i.e.\/}~the
only components of $T_{\mu\nu}$ (where we include any cosmological
constant term in $T_{\mu\nu}$) that change are
\EQ{
\Delta T_{ab}=T_{\phantom{(0)}a}^{(0)\ c}\Delta\tilde g_{cb}\ .
\label{trst}
}
The proof of \eqref{trst} requires a little work and depends upon what kind of
sources are being considered. The first situation which is relevant
to the original Randall-Sundrum as well as the intersecting brane
scenarios, is when the sources are static external sources. In
these cases it is a simple matter to show that the stress-tensor,
including the cosmological constant term, is linear in components of
the background metric: in these cases \eqref{trst} follows
immediately. The second case that is simple to consider is when the
branes are produced by a scalar field with the Lagrangian
\EQ{
{\cal L}= \sqrt{g}
\big[\tfrac12\partial_\mu\phi\partial^\mu\phi
-{\cal V}(\phi )\big]. 
}
In this case, following the
discussion in \cite{US} it is also straightforward to show that the
stress-tensor is linear in the {\it four-dimensional\/} components of
the metric. The point is that $\phi=\phi(z)$ only and so the only
components of the stress-tensor that are altered by $\eta_{ab}\to\tilde
g_{ab}(x)$ are $T_{ab}$ and these are linear in $\tilde g_{ab}(x)$:
\EQ{
T_{ab}=-e^{-A(z)}\tilde
g_{ab}(x)\big[\tfrac12\partial_i\phi(z)\partial^i\phi(z)-{\cal V}(\phi)\big]\ .
}
Furthermore, the background scalar field satisfies an
equation-of-motion that does not
depend on the components of the metric $\tilde g_{ab}$ and so the
field itself is not changed by $\eta_{ab}\to\tilde g_{ab}$. 
Hence, in this case also \eqref{trst} is recovered and furthermore the scalar
field background remains unchanged. Finally, we could consider the
case when the brane is produced by more complicated tensor fields; for
instance this is the situation in string theory. After some analysis
one finds a similar picture in this case as for the scalar field; the
point is that $T_{\mu\nu}$ is linear in components of the metric with
coefficients that are in general functions of 
$\det\tilde g_{ab}(x)$. When Minkowski
space is replaced by the Schwarzschild solution $\det\tilde g_{ab}(x)$
is unchanged and so the change in the stress tensor is linear as in \eqref{trst}.
Furthermore, the equations-of-motion of the tensor field are also not
modified and so the tensor field background remains unchanged \cite{UNP}.
Hence we have established our goal in a very general setting; 
namely, we may replace the background
\eqref{metric} by \eqref{emsm} and still solve Einstein's equations with the same
background fields. 

We now wish to examine the causal structure of the black $p$-brane metric
\eqref{emsm}. Our aim is to show that in the background \eqref{emsm},
there are generically singularities in the transverse space, 
in addition to the usual singularity of the black
hole itself. In order to show this we have to show that (i) these
singularities can be reached by a freely falling observer in finite
proper time and (ii) that the tidal forces experienced by such an
observer will diverge; that is, for each
causal geodesic we should work out the frame components of the Weyl tensor
which are calculated relative to a frame which is parallelly propagated
along the geodesic. Our philosophy is that once we have demonstrated
the existence of these singularities then this indicates that the 
spacetime \eqref{emsm} is pathological and that some other metric
should describe the Schwarzschild solution on the brane. We will then
proceed to intuit the properties of this solution.

Before we calculate the geodesics in detail, however, 
we can glean some information from the curvature
invariants of the spacetime. Indeed, it is
easy to show that the square of the Riemann tensor includes the term
\EQ{
R_{\mu\nu\rho\sigma}R^{\mu\nu\rho\sigma}=R^{(4)}_{abcd}R^{(4)abcd}e^{2A}
+\cdots=\frac{48M^2}{r^6}e^{2A}+\cdots\ ,
\label{cinv}
}
where the ellipsis stands for terms with fewer powers of $e^{A(z)}$. 
In a generic scenario, we expect that the warp-factor $e^{A(z)}\to
\infty$ in some regions in the transverse space.
For instance, in the intersecting-brane case, just
as in the original Randall-Sundrum picture, this is indeed the case:
from \eqref{mdrsmetric} 
the warp factor $e^{A(z)}$ goes to infinity at the ``horizon'' of
AdS.\footnote{The horizon of the patched AdS space is at $\sum_{i=1}^p|z^i|=\infty$.}
Hence the curvature invariant \eqref{cinv} will generically diverge as
we approach the horizon of AdS and also as we approach the 
singular core of the black $p$-brane ($r=0$).  Thus, we suspect
that inertial observers will see infinite tidal forces as they 
approach these regions. In order to confirm this suspicion, we turn to
an analysis of the space of causal geodesics.

To begin, let us consider geodesics in the general background \eqref{gmetric}. At this
stage we do not have to specify a form for the four-dimensional metric
$\tilde g_{ab}$. The geodesic equation
is
\EQ{
\ddot x^\mu+\Gamma^\mu_{\nu\rho}\dot x^\nu\dot x^\rho=0\ ,
\label{geq}
}
where the derivatives are with respect to the affine parameter $\tau$
(taken to be the proper time for time-like geodesics).
These equations imply 
\EQ{
\frac{d}{d\tau}\big(g_{\mu\nu}\dot x^\mu\dot x^\nu\big)=0\ ,
}
or on integrating
\EQ{
g_{\mu\nu}\dot x^\mu\dot x^\nu\equiv e^{-A}\tilde g_{ab}\dot x^a\dot
x^b+g_{ij}\dot z^i\dot z^j=-\sigma\ ,
\label{igeq}
}
where we can take $\sigma=1$ for a time-like geodesic, and $\sigma=0$ for a null
geodesic.

Using the form \eqref{gmetric} for the metric we find that \eqref{geq} for $\mu\equiv
i$ yields
\EQ{
\ddot z^i+\Gamma^i_{jk}\dot z^j\dot z^k-\tfrac12
g^{il}(\partial_le^{-A})\tilde g_{ab}\dot x^a\dot x^b=0\ ,
}
which can be simplified by using \eqref{igeq}, to an equation
involving only the $\{z^i\}$ coordinates:
\EQ{
\ddot z^i+\Gamma^i_{jk}\dot z^j\dot z^k-\tfrac12
g^{il}\partial_lA\big(\sigma+g_{jk}\dot z^j\dot z^k\big)=0\ .
}
We now contract this equation with $g_{il}\dot z^l$ to arrive at
\EQ{
\frac d{d\tau}\Big[e^{-A}\big(\sigma+g_{jk}\dot z^j\dot
z^k\big)\Big]=0\ ,
}
and so
\EQ{
\sigma+g_{jk}\dot z^j\dot z^k=\xi^2 e^A\ ,
\label{zeq}
}
where $\xi$ is a constant of integration. Now we can return to
\eqref{igeq} and use \eqref{zeq} to eliminate the $z^i$-derivatives to
arrive at
\EQ{
\tilde g_{ab}\dot x^a\dot x^b=-\xi^2 e^{2A}\ .
\label{xeeq}
}
This equation has a remarkably simple interpretation. Firstly, let us
define a new affine parameter $\nu$ via
\EQ{
\frac{d\nu}{d\tau}=e^A\ .
\label{apwf}
}
Then \eqref{xeeq} is simply
\EQ{
\tilde g_{ab}\frac{dx^a}{d\nu}\frac{dx^b}{d\nu}=-\xi^2\ ,
\label{effgeq}
}
{\it i.e.\/}~the equation for a
four-dimensional time-like geodesic in the metric $\tilde g_{ab}$ with an
affine parameter related to the higher-dimensional one by the warp
factor \eqref{apwf}. It should not be surprising that a null geodesic
in $(p+4)$-dimensions is equivalent to a time-like geodesic in
four-dimensions: the non-trivial motion in the transverse dimensions
gives rise to a mass in four dimensions. What is gratifying is the
simple relation between the four- and higher-dimensional affine parameters 
\eqref{apwf}.

For the intersecting brane scenario, we can be more specific about the
geodesics. In a given patch 
define $z=\sum_{i=1}^p|z^i|/\sqrt{p}$. The metric \eqref{mdrsmetric}
in this patch is then 
\EQ{
ds^2=\frac1{(1+kz)^2}\big(g_{ab}dx^a\,dx^b+dz^2+dz^i_\perp\,dz^i_\perp\big) 
\ ,
}
where $z^i_\perp$ represent the $p-1$ remaining transverse
coordinates. It is straightforward to find the geodesics in this
case. The null geodesics are straight lines:
\EQ{
z=-c/(k\tau)-1/k\ ,\qquad z^i_\perp=-c^i_\perp/(k\tau)+a_\perp^i\ ,
}
for constants $c$, $c^i_\perp$ and $a^i_\perp$. The time-like geodesics
are, correspondingly, curved:
\EQ{
z=-c/\sin(k\tau)-1/k\ ,\qquad
z^i_\perp=-c^i_\perp/\tan(k\tau)+a_\perp^i\ .
\label{tlg}
}
The affine parameter has been defined in both cases so that
the geodesics approach the horizon $z=\infty$ as $\tau\to 0^-$. Notice that
the geodesics reach the horizon of the AdS space, $z=\infty$, 
after a {\it finite\/} elapse of affine parameter. The
four-dimensional motion of the geodesics is now determined by
\eqref{xeeq} with
\EQ{
\xi^2=\frac{c^2+c_\perp^ic_\perp^i}{k^2c^4}\ .
}

Returning to the more general case, we now take 
the four-dimensional metric $\tilde g_{ab}$ to be the 
Schwarzschild metric of a black hole. In that case we can 
specify the behaviour of the geodesics in four dimensions. Firstly,
the metric \eqref{emsm} has two Killing vectors: $k=\partial/\partial
t$ and $m=\partial/\partial\phi$ which give rise to the conserved
quantities $E=-k\cdot u$ and $L=m\cdot u$, where $u=\dot x^\mu$ is the
velocity along a geodesic. Rearranging gives
\EQ{
\dot t=\frac{Ee^A}U\ ,\qquad \dot\phi=\frac{Le^A}{r^2\sin^2\theta}\ .
}
Since there are two conserved quantities and we may consider 
motion in the equatorial plane $\theta=\pi/2$, the effective equation for radial
motion may be deduced from \eqref{effgeq}:
\EQ{
\dot r^2+e^{2A}\Big[\Big(\xi^2+\frac{L^2}{r^2}\Big)U(r)-E^2\Big]=0\ .
}
By introducing the new affine parameter $\nu$ in \eqref{apwf} and
rescaling $\tilde r=r/\xi$, $\tilde E=E/\xi$, $\tilde L=L/\xi^2$ and
$\tilde M=M/\xi$, this can be written as
\EQ{
\Big(\frac{d\tilde r}{d\nu}\Big)^2+\Big(1+\frac{\tilde L^2}{\tilde
r^2}\Big)\Big(1-\frac{2\tilde M}{\tilde r}\Big)=\tilde E^2\ ,
}
which is precisely the radial geodesic equation for a four-dimensional
Schwarzschild black hole of mass $\tilde M$. Note that $\nu$ is the
proper time along this four-dimensional geodesic.

{}From here, the analysis is very similar to that performed in
\cite{CHR}. To wit, there are two distinct classes of time-like geodesics
which experience infinite affine parameter $\nu$: those which are
bound states (ones that orbit in a restricted finite range of
${\tilde r}$), and those which are not (ones which make it to
${\tilde r} = \infty$).
For the orbits which escape to ${\tilde r} = \infty$ the late time
behaviour is
\EQ{
\tilde r\thicksim\nu\sqrt{\tilde E^2-1}
}
and consequently we recover the integral 
\EQ{
r\thicksim \sqrt{E^2-\xi^2}\int^\tau e^{A}\,d\tau\ .
\label{stgb}
}
In the intersecting-brane scenario, consider a time-like geodesic
\eqref{tlg} with $z_\perp^i=$ constant and so $z=-c/\sin(k\tau)-1/k$. In
this case $\nu=-(1/\xi^2k){\rm cot}(k\tau)$. So along such a geodesic, which
is a bound state in four-dimensions (so that $r$ remains finite) 
it is easy to see that the curvature invariant \eqref{cinv} diverges
at the horizon of AdS. 
So for such geodesics there is a genuine singularity there. However, 
for the second type of geodesics which escape to $r=\infty$ we have from \eqref{stgb}
$r\propto{\rm cot}(k\tau)$ and so along these geodesics the curvature
invariant \eqref{cinv} remains finite. In order to establish the
existence of a singularity at the horizon along these geodesics, 
we should examine the frame components of the
Riemann tensor in an orthonormal frame which has been parallelly
propagated along the geodesic. These frame components will
measure the tidal forces experienced by the free-falling observer who
moves along the geodesic.

We may calculate that the tangent vector to such a non-bound state
geodesic (for $L=0$) is given as\footnote{Where we are using the
$(t,r,\theta,\phi,z^i)$ ordering for the components.} 
\EQ{
u^\mu=\Big(e^AE/U(r),e^A\sqrt{E^2-\xi^2 U(r)},0,0,\dot z^i\Big)\ .
}
A parallelly propagated unit normal to the geodesic is likewise given as
\EQ{
n^\mu=\Big(e^{A/2}\sqrt{E^2-\xi^2
U(r)}/(\xi U(r)),e^{A/2}E/\xi,0,0,0\Big)\ .
}
Using just these two orthonormal vectors we can see a potential divergent tidal force.
Indeed, one of the frame components is calculated to be
\EQ{
R_{(u)(n)(u)(n)} = R_{\mu\nu\rho\sigma}u^\mu n^\nu u^\rho
n^\sigma=-\frac{2M\xi^2}{r^3}e^{2A}+\cdots\ ,
}
where the ellipsis represents less singular terms.
In the intersecting-brane scenario this behaves as $1/(\sin^4(k\tau){\rm
cot}(k\tau)^3)$ which {\it does\/} diverge at the AdS horizon.
It follows that the black $p$-brane is singular all the way along the
AdS horizon.  

It is well known that black strings and $p$-branes in asymptotically flat
space are unstable to long-wavelength perturbations---the 
``Gregory-Laflamme instability'' \cite{GREGORY}.
A black hole horizon is entropically preferred to a sufficiently large ``patch''
of $p$-brane horizon. Thus, a black $p$-brane horizon will generically want to 
fragment and form an array of black holes. The argument is worth
recalling. The relevant situation to consider in the present context
is a four-dimensional Schwarzschild black hole embedded in flat
$(p+4)$-dimensional spacetime; {\it i.e.\/}~a $p$-brane in $p+4$ dimensions. 
Let $R_4$ be the radius of the horizon of the black $p$-brane which is
related to the associated four-dimensional
Schwarzschild mass of the solution by $R_4=2G_4M_4$.
Hence the entropy for a portion of
such an object with ``area'' $L^p$, in the $p$-dimensional
transverse space, is $\sim L^pR_4^2$. This object has
an effective $(p+4)$-dimensional mass of $M_*=L^pR_4/G_*$, where $G_*$
is the $(p+4)$-dimensional Newton constant. Let us compare this to a
a $(p+4)$-dimensional black hole carrying the same
mass. Such an object would have a horizon radius
$R_*=2(G_*M_*)^{1/(p+1)}$ and hence an entropy $\sim
R_*^{p+2}=(G_*M_*)^{(p+2)/(p+1)}=(L^pR_4)^{(p+2)/(p+1)}$. 
So when $(L^pR_4)^{(p+2)/(p+1)}\simeq
L^pR_4^2$, {\it i.e.\/}~$L\simeq R_4$, 
we expect that the black $p$-brane becomes unstable with
respect to the hyperspherical black hole. Another way to say this is
that there will be a destabilizing mode of wavelength $\sim L$, {\it
i.e.\/}~with a wavelength $\lambda\sim R_4$.

One might suspect that a similar instability occurs for black $p$-branes in
spacetimes that are asymptotically AdS.  In \cite{CHR} the authors
argued that a black string (or 1-brane) in AdS will generically have to pinch off
down near the AdS horizon.  This is because at large $z$ the string is
so ``skinny'' that it does not see the curvature scale of the ambient AdS
space, and so the argument of Gregory and Laflamme goes through as it
would for a string in flat space. Generalizing to the
intersecting-brane scenario, the $p$-brane is very thin at large $z$
because the proper radius of its horizon gets warped:
\EQ{
R_4(z)=e^{-A(z)/2}R_4=\frac{R_4}{1+kz}\ .
}
Hence using the logic of the
Gregory-Laflamme instability, at a given $z$ the $p$-brane is unstable
to a mode of wavelength $\lambda(z)\sim R_4(z)$. 
The important point is that AdS acts like a box of size $\sim k^{-1}$
and so can only allow unstable perturbations of wavelength $\lesssim
k^{-1}$. Hence, there exists a critical value for the warp-factor 
when the wavelength of the destabilizing mode can just fit inside the AdS
box:
\EQ{
e^{-A(z_{\rm crit})/2}R_4\simeq k^{-1}\ ,
}
or in this case 
\EQ{
z_{\rm crit}\simeq R_4-k^{-1}\ .
\label{zcrit}
}
This corresponds to a proper distance
\EQ{
r_{\rm crit}\simeq k^{-1}\log(kR_4)\equiv k^{-1}\log(2kG_4M_4)\ ,   
}
from the junction. 
Thus, at sufficiently large $z$ any large perturbation
will fit inside the AdS ``box'', and so an instability will occur near
the AdS horizon.  On the other hand, when $z$
is small enough the potential instability occurs at wavelengths much
larger than $k^{-1}$ and so the instability will not occur in this region.
Just as for the black string in AdS, a black $p$-brane in AdS is unstable
near the AdS horizon but stable far from it.

After the black $p$-brane fragments, a stable portion of horizon will
remain ``tethered'' to the boundary of AdS.  Of course,
if we are in the intersecting brane scenario then the boundary of
AdS has been cut away and so this stable remnant of horizon 
will envelop the brane junction.  The detached pieces of horizon will
presumably fall into the bulk of AdS.\footnote{Since we do not
actually know the explicit form of a metric which can describe the dynamics
of such a situation, we have to use
our intuition here.} The stable remnant of $p$-brane horizon acts as if
it has a tension, balancing the force pulling it towards the center of
AdS.
This remnant portion of horizon, far from being a spherically symmetric
black hole, will be a highly deformed black object in $(p+4)$ dimensions.
It is amusing to think about the gross properties of this object.
In \cite{CHR}, the authors argued that after the black string fragments,
the stable object left behind would resemble a ``black cigar'' (or 
more realistically a ``pancake'' because $R_4\gg k^{-1}$
implying that the object only extends a small proper distance
in the transverse space compared with the brane~\cite{EHM,GKR}).

When $p=2$, we see that the black membrane will tend to fragment at some
surface where $z=(|z^1|+|z^2|)/\sqrt{2}=$ constant given by 
\eqref{zcrit}.  In other words, the
black 2-brane will tend to fragment along a diamond shaped 
surface.  After fragmentation, the force balance between the horizon tension
and the AdS potential and the symmetry of the problem will preserve this basic shape.
In other words, a black hole at the junction of two domain walls in 
$AdS_6$, far from being spherically symmetric, will be deformed into
the shape of a ``black diamond'' as illustrated in Figure 1.
In general, for an arbitrary number $p$ domain walls, the horizon will be deformed
towards the shape of a polyhedron with $2^p$ sides.  Each side will
be a portion of horizon corresponding to a given ``patch'' of $AdS_{p+4}$ used
to construct the domain wall junction.  
\begin{figure}
\begin{center}
\includegraphics[height=10cm]{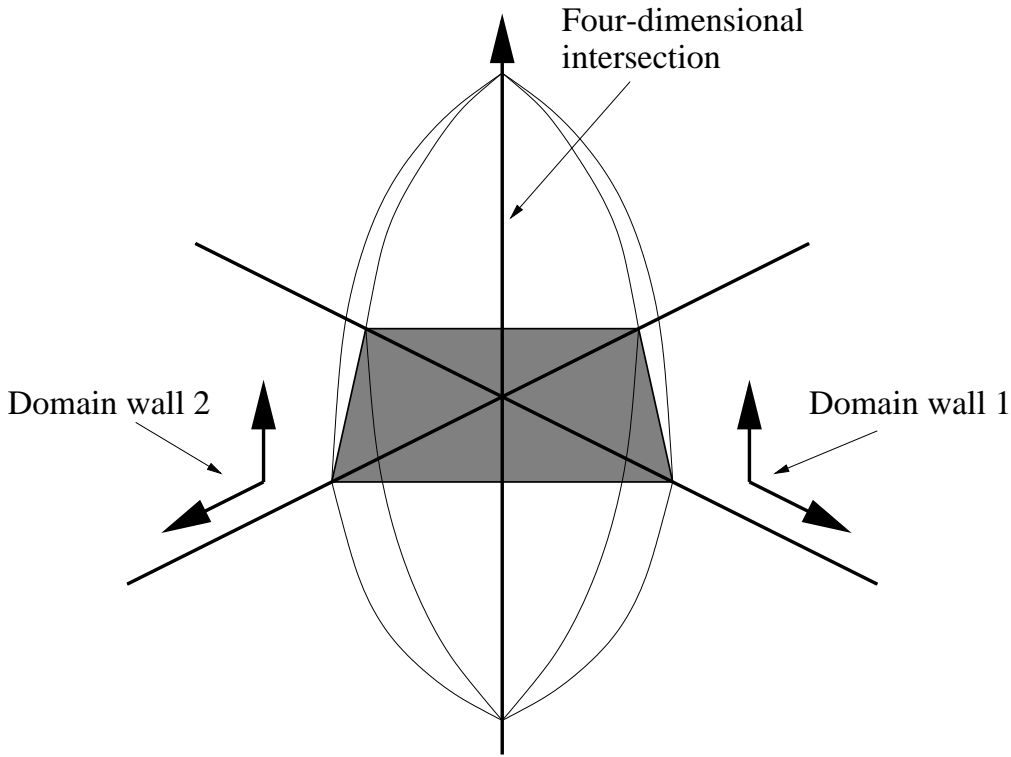}
\end{center}
\caption{\small The event horizon of a brane-world black hole in a scenario where the brane-world
is the junction between two domain walls in six-dimensional anti-de Sitter
space. Any cross-section of the horizon perpendicular to the
brane-world has a ``black diamond'' profile.}
\label{fig:potential}
\end{figure}

In this paper we have studied aspects of gravitational collapse
in certain brane-world scenarios, where gravity is
localized on some four-dimensional submanifold of a higher dimensional
space. If this scenario is to be
phenomenologically viable, then a brane-world observer should see
that the endpoint of the gravitational collapse of
uncharged non-rotating matter trapped on the brane is, at least to a
good approximation, the Schwarzschild solution. In other words,
there should exist a metric
in the higher-dimensional bulk spacetime which induces an
approximation to the Schwarzschild solution on the brane up to
corrections that are small for $r\gg k^{-1}$.
We have shown that when one intersects the four-dimensional world with 
a black $p$-brane, the induced metric at the junction is
exactly the Schwarzschild solution. However, in a concrete example
corresponding to the intersecting branes, we  have analyzed the causal structure of
this solution, and found that the AdS horizon is singular.  In fact, we
found that the horizon region is a ``pp-curvature'' singularity
\cite{Chamblin}, which simply means that parallelly propagated frame
components of the curvature tensor diverge as the region is approached along
causal geodesics, whereas scalar curvature invariants do not necessarily
diverge.  This singularity will be visible from the brane-world, and one
might regard this as a pathology of the model.\footnote{However, as in \cite{Chamblin},
we would argue that anything emerging from the singularity at the AdS
horizon will be heavily red-shifted by the time it reaches the brane,
and therefore it will likely be heavily suppressed.}
At any rate, we have argued that the black $p$-brane solution is unstable,
and that the brane horizon will want to fragment near the AdS horizon.  Presumably,
the portions of the brane horizon which break away from the brane-world might fall
into the bulk of AdS and form a bulk black hole.  We have suggested that at
late times this system settles down to a deformed horizon which intersects the
brane-world in such a way that the metric induced at the domain wall junction
will still be approximately the Schwarzschild solution.  While we do not know the exact metric
describing this configuration, we conjecture that this metric exists and that
it is the {\it unique} stable vacuum solution that describes a non-rotating
uncharged black hole on the domain wall junction.

Our analysis can easily be extended to other cases. For instance, we may
 consider smoothed versions of the Randall-Sundrum scenario where
the domain wall is created by some matter field
\cite{cveticdw,DFGK,Chamblin,Gremm,US}. We have shown that in this case also the brane can be
intersected by a black string without changing the matter field background.
In this case, far from the brane the geometry
approaches that of $AdS_5$ and so the analysis of the singularities
at the AdS horizon that we have presented is equally
valid in this case. We can also easily apply our analysis to
higher-dimensional cases that do not correspond to intersecting branes;
for instance three-branes embedded in dimension $d>5$ \cite{CK,Erich,US}.

Finally, it would clearly be desirable to find the exact form of the
metric describing the end-point of gravitational collapse on the brane junction
and hence determine the corrections to the Schwarzschild metric.

\acknowledgments

We thank Sean Carroll for a stimulating conversation. A.C. is supported in part 
by funds provided by the U.S. Department of Energy under
cooperative research agreement DE-FC02-94ER40818.  A.C. thanks the
T-8 group at Los Alamos for hospitality while this project
was initiated.  C.C. is an Oppenheimer Fellow at the 
Los Alamos National Laboratory.
C.C., J.E. and T.J.H. are supported 
by the US Department of Energy under contract W-7405-ENG-36.

\end{document}